\documentclass[a4paper,11pt]{article}
\pdfoutput=1 % if your are submitting a pdflatex (i.e. if you have
\usepackage{jheppub} % for details on the use of the package, please
\usepackage[T1]{fontenc} % if needed

\usepackage{graphicx}
\usepackage{amssymb, amsmath,mathtools,amsthm}

\newcommand{\be}{\begin{equation}}
\newcommand{\bea}{\begin{align}}
\newcommand{\eea}{\end{align}}
\newcommand{\beq}{\begin{equation}}
\newcommand{\ee}{\end{equation}}
\newcommand{\eeq}{\end{equation}}

\def\ip{${\cal I}^+$}
\def\im{${\cal I}^-$}
\def\hp{${\cal H}^+$}
\def\hm{${\cal H}^-$}
\def\ipp{${\cal I}^+_+$}

\def\imm{${\cal I}^-_-$}
\def\hpp{${\cal H}^+_+$}

\def\h{${\cal H}$}
\def\i{${\cal I}$}
\def\zb{{\bar{z}}}

\title{\boldmath BMS invariance and the membrane paradigm}

\author[a]{Robert F. Penna}

\affiliation[a]{Department of Physics and Kavli Institute for Astrophysics and Space Research,\\
Massachusetts Institute of Technology, Cambridge, Massachusetts 02139, USA}

\emailAdd{rpenna@mit.edu}

\abstract{

The Bondi-van der Burg-Metzner-Sachs (BMS) group is the asymptotic symmetry group of asymptotically flat spacetime.  It is infinite dimensional and entails an infinite number of conservation laws.  According to the black hole membrane paradigm, null infinity (in asymptotically flat spacetime) and black hole event horizons behave like fluid membranes.  The fluid dynamics of the membrane is governed by an infinite set of symmetries and conservation laws.  Our main result is to point out that the infinite set of symmetries and conserved charges of the BMS group and the membrane paradigm are the same.  This relationship has several consequences.  First, it sheds light on the physical interpretation of BMS conservation laws.  Second, it generalizes the BMS conservation laws to arbitrary subregions of arbitrary null surfaces.  Third, it clarifies the identification of the superrotation subgroup of the BMS group.  We briefly comment on the black hole information problem.

}

\begin{document} 
\maketitle
\flushbottom

\section{Introduction}
\label{sec:intro}

The Bondi-van der Burg-Metzner-Sachs (BMS) group is the asymptotic symmetry group of asymptotically flat spacetime \cite{1962RSPSA.269...21B,1962RSPSA.270..103S,deBoer:2003vf,2010PhRvL.105k1103B,Barnich:2010eb,Barnich:2011ct,2014JHEP...07..152S,2014PhRvD..90l4028C,2014arXiv1404.4091C,2015JHEP...04..076C,2015arXiv150207644K}.  It is infinite dimensional and entails an infinite number of conservation laws.  These have been interpreted as ``energy and momentum conservation at every angle'' \cite{2014JHEP...07..152S,2015arXiv150206120P}.  The BMS group is related to soft theorems and the gravitational memory effect \cite{2014arXiv1411.5745S,2015arXiv150206120P}.  It may be relevant for the black hole information problem \cite{Hawking:1976ra,'tHooft:1995ij,'tHooft:1996tq,Polchinski:2015cea,Hooft:2015jea,Hawking:2016msc}.  It has the same relationship to asymptotically flat spacetimes as the conformal group has to anti de Sitter spacetimes, so it can be expected to govern holographic descriptions of asymptotically flat spacetimes.

This paper describes a relationship between the BMS group and the black hole membrane paradigm.  According to the membrane paradigm, black hole event horizons behave like 2+1 dimensional fluids \cite{Damour:1979wya,1982mgm..conf..587D,Price:1986yy,1986bhmp.book.....T,1998PhRvD..58f4011P}.   Future null infinity in asymptotically flat spacetimes can also be described as a 2+1 dimensional fluid \cite{2015PhRvD..91h4044P}.  The fluid dynamics of the membrane is governed (as we will show) by an infinite set of symmetries and conserved charges.  Our main result is to point out that these symmetries and charges are the same as the BMS symmetries and charges.  This gives a new perspective on the BMS group and the membrane paradigm.

This new perspective has a number of advantages. First, it clarifies the physical interpretation of BMS conservation laws.  Conservation of energy and momentum at every angle are equivalent to the energy conservation equation and the Damour-Navier-Stokes equation governing the membrane.

Second, the membrane paradigm gives a generalization of the usual BMS conservation laws to arbitrary subregions of arbitrary null surfaces. The generalized BMS conservation laws can be applied to the event horizons of one-sided black holes formed from stellar collapse and de Sitter horizons.  The infinite set of conserved charges can be computed directly from the fluid stress-energy tensor, without first finding asymptotic fall-off conditions for the metric or computing asymptotic Killing vectors.  

Third, the membrane paradigm clarifies the nature of the superrotation subgroup of the BMS group.  The superrotation subgroup has been variously identified with the set of Lorentz transformations \cite{1962RSPSA.269...21B,1962RSPSA.270..103S}, the set of infinitesimal local conformal transformations \cite{2010PhRvL.105k1103B,Barnich:2011ct}, and the diffeomorphism group of the sphere, Diff($S^2$)  \cite{2014PhRvD..90l4028C,2015JHEP...04..076C}.  The membrane paradigm suggests the superrotation subgroup is Diff($S^2$).

The connection between the BMS group and the membrane paradigm suggests the membrane should be taken seriously, as physical degrees of freedom living on null surfaces.  Related suggestions have been made by \cite{Balachandran1996581}.  Note however that the membrane's degrees of freedom are somewhat observer-dependent.  Each observer assigns the membrane degrees of freedom to the boundaries of their causal diamond.   Observers in the exterior of an asymptotically flat black hole see membranes on the event horizon and future null infinity.  Observers in the black hole interior each have different causal diamonds in general.

This paper is organized as follows.  Section \ref{sec:membrane} reviews the membrane paradigm, section \ref{sec:stationary} discusses stationary spacetimes, section \ref{sec:nonstationary} discusses nonstationary spacetimes, and section \ref{sec:Q} discusses the extension to electrodynamics.

\section{The membrane paradigm}
\label{sec:membrane}

The membrane paradigm attaches a 2+1 dimensional fluid stress-energy tensor to null surfaces.  Observers in the exterior of an asymptotically flat black hole assign membranes to the event horizon and future null infinity.  In this section, we review the definition of the membrane stress-energy tensor.  For an explanation of why this definition is correct, see \cite{Price:1986yy,1986bhmp.book.....T,1998PhRvD..58f4011P,2015PhRvD..91h4044P}.  The stress-energy tensor is an integral over the extrinsic curvature of the membrane, so we first define extrinsic curvature.

The membrane is a timelike cutoff surface placed slightly outside the null surface.  At the event horizon, the cutoff surface is the stretched horizon.  At future null infinity, the cutoff surface is ``stretched infinity,'' a large but finite sphere.  Let $n$ be the unit normal of the membrane.  The projection tensor,
\beq
h_{ab} = g_{ab} - n_a n_b,
\eeq
is the metric induced on the membrane by the 3+1 dimensional spacetime metric, $g_{ab}$.  Let $U^a$ be the world lines of a family of fiducial observers.  The metric on constant-time slices of the membrane is
\beq
\gamma_{ab}=h_{ab} +U_a U_b.
\eeq
We use Greek indices $\mu,\nu,\dots$ for tensors on 3+1 dimensional spacetime, lower case roman indices $a,b,\dots$ for tensors on the 2+1 dimensional membrane, and upper case roman indices $A,B,\dots$ for tensors on constant-time slices of the membrane.  We denote the 4-covariant derivative by $\nabla_\mu$, the 3-covariant derivative by $_{|a}$, and the 2-covariant derivative by $_{\parallel A}$.

The extrinsic curvature of the membrane is
\beq\label{eq:K}
K^a_b = h^c_b \nabla_c n^a,
\eeq
where we follow the sign convention of \cite{1998PhRvD..58f4011P} rather than \cite{Price:1986yy,1986bhmp.book.....T,2015PhRvD..91h4044P}, which differs by an overall minus sign.  

The membrane's stress-energy tensor is
\beq\label{eq:t}
t_{ab} = \pm \frac{1}{8\pi}\left(K h_{ab} - K_{ab}\right),
\eeq
where the upper sign applies at event horizons and the lower sign applies at null infinity.  The sign difference comes from the fact that the membrane terminates the gravitational field and the horizon is an inner boundary of spacetime while infinity is an outer boundary.  See \cite{2015PhRvD..91h4044P} for details.  The energy density is
\beq\label{eq:Sigma}
\Sigma = t_{ab} U^a U^b = -\frac{\theta}{8\pi},
\eeq
where $\theta$ is the expansion scalar (defined below).  The momentum density is 
\beq\label{eq:pi}
\pi_A = t_{aA} U^a,
\eeq
and the stress tensor is
\beq
t_{AB}  = p\gamma_{AB} -2\eta \sigma_{AB}-\zeta\theta\gamma_{AB}.
\eeq
$\eta$ and $\zeta$ are the shear and bulk viscosity coefficients, respectively, and $\sigma_{AB}$ is the shear tensor. The pressure is
\beq\label{eq:p}
p=\frac{g}{8\pi},
\eeq
where $g$ is the surface gravity of the membrane.

The expansion and shear are
\begin{align}
\theta &= \pm K^A_A,\label{eq:theta}\\
\sigma_{AB} &= \pm K_{AB}-\frac{1}{2}\theta \gamma_{AB},\label{eq:sigma}
\end{align}
where upper signs apply at future event horizons and past null infinity, while lower signs apply at past event horizons and future null infinity.  The choice of sign depends on whether the membrane satisfies ingoing (upper signs) or outgoing (lower signs) boundary conditions \cite{2015PhRvD..91h4044P}.  The membrane has vanishing rotation,
\beq\label{eq:omega}
\omega_{AB} = \pm K_{[AB]}=0,
\eeq
because $n_a$ is hypersurface orthogonal.  

The viscosity coefficients are
\beq\label{eq:viscp}
\eta_+ = 1/(16\pi), \quad \zeta_+ = -1/(16\pi)
\eeq
on future event horizons and future null infinity, and
\beq\label{eq:viscm}
\eta_- = -1/(16\pi), \quad \zeta_- = 1/(16\pi)
\eeq
on past event horizons and past null infinity \cite{2015PhRvD..91h4044P}.

\section{Stationary spacetimes}
\label{sec:stationary}

The Einstein equations imply that the membrane obeys the Damour-Navier-Stokes equation  \cite{Damour:1979wya,1982mgm..conf..587D,Price:1986yy,1986bhmp.book.....T,1998PhRvD..58f4011P,2015PhRvD..91h4044P},
\beq\label{eq:ns}
\mathcal{L}_U \pi_A + \nabla_A p - \zeta \nabla_A \theta - 2 \eta \sigma^B_{A\parallel B}+T^M_{nA}=0,
\eeq
where $T^M_{nA}$ represents non-gravitational sources of momentum.   In stationary spacetimes, it is possible to choose a slicing for which $\theta=\sigma_{AB}=0$ and $p$ is constant \cite{Price:1986yy}.  We will further assume $T^M_{nA}=0$ for the remainder of this section. In this case, eq. \eqref{eq:ns} is simply
\beq
\mathcal{L}_U \pi_A = 0,
\eeq
and the momentum density, $\pi_A$, is conserved.  This implies an infinite set of conserved charges,
\beq\label{eq:Q}
Q_{f,Y^A}=\int d^2x \sqrt{\gamma} (f p - Y^A \pi_A),
\eeq
where $f$ and $Y^A$ are arbitrary functions and the integral is over constant-time slices of the membrane.   
Setting $f=\delta^2(x^P-\hat{x}^P)$ and $Y^A=0$ gives a set of charges corresponding to ``energy at every angle.'' Setting $f=0$ and $Y^A=\delta^2(x^P-\hat{x}^P) \delta^A_B$ gives a set of charges corresponding to ``momentum at every angle.''

There is some arbitrariness in the normalization of the charges \eqref{eq:Q}.  For example, it would be equally natural to choose
\beq\label{eq:Qp}
Q'_{f,Y^A}	= \int d^2x \sqrt{\gamma} (f t^a_a - Y^A \pi_A) 
		= \int d^2x \sqrt{\gamma} (2 f p - Y^A \pi_A),
\eeq
where $t^a_a=2p$ is the trace of the membrane's stress-energy tensor.

These charges can be computed for any null surface.  In particular, they can be computed at event horizons and at future null infinity (in asymptotically flat spacetimes) .  In the next two subsections, we check that the charges so defined are the same as the BMS charges.

\subsection{Event horizons}

The near horizon geometry of a stationary 3+1 dimensional black hole can be expressed as \cite{Donnay:2015abr}
\beq\label{eq:dsH}
ds^2 = f dv^2 +2kdvdp + 2g_{vA}dvdx^A + g_{AB}dx^A dx^B,
\eeq
where $\rho\rightarrow 0$ at the horizon and $x^A$ are the angular directions on the horizon.  In the near horizon limit,
\begin{align}
f &=-2\kappa p+\mathcal{O}(\rho^2),\\
k &=1+\mathcal{O}(\rho^2),\\
g_{vA} &= \rho \theta_A + \mathcal{O}(\rho^2),\\
g_{AB} &= \Omega m_{AB}+\rho \lambda_{AB}+\mathcal{O}(\rho^2),
\end{align}
where $\theta_A$ and $\Omega$ are functions of $x^A$ only, $\lambda^{AB}=\lambda^{AB}(v,x^A)$, and $m_{AB}$ is the metric of the unit two-sphere.  The components $g_{\rho A}$ and $g_{\rho\rho}$ are $\mathcal{O}(\rho^2)$.  The surface gravity of the horizon is $\kappa$ and the pressure \eqref{eq:p} is
\beq
p=\frac{\kappa}{8\pi}.
\eeq

The membrane's unit normal is
\beq
n=\alpha^{-1} dr,
\eeq
where $\alpha=\sqrt{2\kappa\rho}$ is the lapse.  Plugging into \eqref{eq:K} gives the extrinsic curvature of the membrane and using \eqref{eq:t} gives the stress-energy tensor.  The momentum density \eqref{eq:pi} is 
\beq
\pi_A = \alpha t^v_A = \frac{1}{16\pi} \theta_A + \mathcal{O}(\rho).
\eeq
The conserved charges \eqref{eq:Q} are 
\beq
Q_{f,Y^A}=\frac{1}{16\pi}\int d^2x \sqrt{m} \Omega (2f \kappa - Y^A \theta_A),
\eeq
This infinite set of charges precisely coincides with the BMS charges at the horizon \cite{Donnay:2015abr}, except that \cite{Donnay:2015abr} assume $Y^A$ is a conformal Killing vector.  This restriction appears artificial from the perspective of the membrane paradigm.  We return to this point in Sec. \ref{sec:super}.

\subsection{Null infinity}

Consider a stationary, 3+1 dimensional asymptotically flat spacetime.  The metric near null infinity may be put into the form \cite{Flanagan:2015pxa}
\beq\label{eq:dsinf}
ds^2 = -U du^2 -2dudr + 2g_{uA}dudx^A + r^2m_{AB}dx^A dx^B,
\eeq
where 
\begin{align}
U(r) &=  1-\frac{2M}{r}+\mathcal{O}(r^{-2}),\\
g_{uA} &= \frac{2}{3}\frac{N_A}{r}+\mathcal{O}(r^{-2}).
\end{align}
$M$ is a constant and $N_A$ is a function of $x^A$ only.

The membrane's unit normal is
\beq
n=\alpha^{-1} dr,
\eeq
and $\alpha=\sqrt{1-2M/r}$ is the lapse.  The surface gravity is 
\beq
\kappa = \frac{M}{r^2}.
\eeq
This vanishes as $r\rightarrow \infty$, but $\sqrt{\gamma}\kappa$ is finite, so the first term in the charge \eqref{eq:Q} is finite.  The momentum density \eqref{eq:pi} is 
\beq
\pi_A = \alpha t^u_A = -\frac{N_A}{8\pi r^2} + \mathcal{O}(r^{-3}).
\eeq
Again, this vanishes as $r\rightarrow \infty$, but $\sqrt{\gamma}\pi_A$ gives a finite contribution to the charge \eqref{eq:Q}.  The conserved charges \eqref{eq:Qp} are 
\beq
Q'_{f,Y^A}=\frac{1}{16\pi}\int d^2x \sqrt{m}(4fM + 2Y^A N_A),
\eeq
This set of charges is the same as the infinite set of BMS charges at null infinity (for stationary, asymptotically flat spacetimes) \cite{2011JHEP...12..105B,Flanagan:2015pxa}, except that again there is some disagreement about the allowable $Y^A$.  

\subsection{Superrotations}
\label{sec:super}

The charges generated by $Y^A$ correspond to the superrotation charges of the BMS group.  From the perspective of the membrane paradigm, $Y^A$ can be any smooth vector field on the sphere.  The set of all smooth vector fields on the sphere is the Lie algebra of Diff($S^2$), the diffeomorphism group of the sphere.  So we identify the superrotation group with Diff($S^2$).

The identification of the superrotation group has generated confusion.  The superrotation group was originally identified with the Lorentz group \cite{1962RSPSA.269...21B,1962RSPSA.270..103S}.  It has recently been suggested that the superrotation group should be extended to include infinitesimal local conformal transformations \cite{2010PhRvL.105k1103B,Barnich:2011ct}.   It has also been suggested that the superrotations should be identified with Diff($S^2$) \cite{2015JHEP...04..076C}.  The argument of \cite{2015JHEP...04..076C} is based on the relationship of the BMS group to soft theorems.  We have used the relationship of the BMS group to the membrane paradigm to give a new reason for identifying the superrotation group with Diff($S^2$).

\subsection{General null surfaces}

One of the advantages of the membrane paradigm is that it immediately extends the definition of BMS charges  to any null surface.  The charges \eqref{eq:Q} can be computed at de Sitter horizons, the event horizons of one-sided black holes formed from stellar collapse and, more generally, the boundaries of any observer's causal diamond.    Charge conservation follows from the Damour-Navier-Stokes equation \eqref{eq:ns} governing the membrane.  The charges \eqref{eq:Q} can be computed directly, without first finding the asymptotic metric or computing asymptotic Killing vectors.

\section{Nonstationary spacetimes}
\label{sec:nonstationary}

The Damour-Navier-Stokes equation governing the membrane is \eqref{eq:ns},
\beq\label{eq:ns2}
\mathcal{N}_A \equiv \mathcal{L}_U \pi_A + \nabla_A p - \zeta \nabla_A \theta - 2 \eta \sigma^B_{A\parallel B} +T^M_{nA}=0.
\eeq
In nonstationary spacetimes, each term can be nonzero.  Contracting \eqref{eq:ns2} with an arbitrary vector field, $Y^A$, and integrating over the entire 2+1 dimensional membrane gives an infinite set of conservation laws:
\beq\label{eq:N}
\int_R dx^3 \sqrt{h} \thinspace Y^A \mathcal{N}_A = -\int_{R^c} dx^3 \sqrt{h} \thinspace Y^A \mathcal{N}_A,
\eeq
where $R$ is any region of the membrane and $R^c$ is its complement.  If the membrane is null infinity, we may take $R=\mathcal{I}^+$ and $R^c=\mathcal{I}^-$ to be future and past null infinity.  If the membrane is the event horizon of an eternal black hole, we may take $R=\mathcal{H}^+$ and $R^c=\mathcal{H}^-$ to be the future and past event horizons.  
If the membrane is the event horizon of a one-sided black hole formed from stellar collapse, we may take $R=\mathcal{H}^+$ and set the rhs of \eqref{eq:N} equal to zero.

The membrane is further governed by the fluid energy equation \cite{Damour:1979wya,1982mgm..conf..587D,Price:1986yy,1986bhmp.book.....T,1998PhRvD..58f4011P,2015PhRvD..91h4044P},
\beq\label{eq:energy}
\mathcal{M}\equiv \mathcal{L}_U \Sigma + \theta \Sigma + p\theta -\zeta \theta^2 - 2 \eta \sigma_{AB}\sigma^{AB}-T^M_{nU}=0,
\eeq
where $T^M_{nU}$ represents nongravitational fluxes of energy into the membrane.   Multiplying this equation by an arbitrary function, $f$, and integrating over the membrane gives another infinite set of conservation laws:
\beq\label{eq:M}
\int_R dx^3 \sqrt{h} f\mathcal{M} = -\int_{R^c} dx^3 \sqrt{h} \thinspace f\mathcal{M}.
\eeq

The conservation laws \eqref{eq:N} and \eqref{eq:M} are more general than the standard BMS conservation laws: they apply to arbitrary subregions of arbitrary null surfaces.  They may be applied to the event horizons of one-sided black holes formed from stellar collapse and to de Sitter horizons.  In the next subsection, we check that \eqref{eq:M} coincides with the standard BMS conservation laws at null infinity. 

\subsection{Null infinity}

Assume $R=\mathcal{I}^+$ and $R^c=\mathcal{I}^-$.   The metric near \ip\ in retarded Bondi coordinates is \cite{1962RSPSA.270..103S,Barnich:2010eb,2014JHEP...07..152S}
\begin{align}\label{eq:metricip}
ds^2	& = -du^2 - 2dudr + 2r^2 \gamma_{z\zb}dzd\zb \notag\\
 	& + \frac{2m_B}{r}du^2 + r C_{zz}dz^2+rC_{\zb\zb}d\zb^2-2U_zdudz-2U_{\zb} dud\zb + \dots,
\end{align}
where 
\begin{align}
\gamma_{z\bar{z}} &= \frac{2}{(1+z\zb)^2}\\
U_z &= -\frac{1}{2}D^z C_{zz},
\end{align}
and $D_z$ is the $\gamma$-covariant derivative.  The Bondi news tensor is
\beq
N_{zz} = \partial_u C_{zz}.
\eeq
The functions $m_B, C_{zz},\dots$ appearing in \eqref{eq:metricip} are functions of $u,z,\zb$ only.  Indices of $C_{zz}, U_z, D_z$ and $N_{zz}$ are raised and lowered with $\gamma_{z\zb}$.

The metric near \im\ in advanced Bondi coordinates is \cite{1962RSPSA.270..103S,Barnich:2010eb,2014JHEP...07..152S}
\begin{align}\label{eq:metricim}
ds^2	& = -dv^2 + 2dvdr + 2r^2 \gamma_{z\zb}dzd\zb \notag\\
 	& + \frac{2m_B^-}{r}dv^2 + r D_{zz}dz^2+rD_{\zb\zb}d\zb^2-2V_zdvdz-2V_{\zb} dvd\zb + \dots,
\end{align}
where 
\beq
V_z = \frac{1}{2}D^z D_{zz},
\eeq
and the news tensor is
\beq
M_{zz} = \partial_v D_{zz}.
\eeq
Near infinity, the usual time coordinate is related to $u$ and $v$ by
\beq
u=t-r, \quad v=t+r.
\eeq

Expanding the membrane energy equation \eqref{eq:energy} near \ip\, using retarded Bondi coordinates \eqref{eq:metricip}, gives
\beq\label{eq:constraint}
\frac{\partial_u m_B + \frac{1}{2}\partial_u \left[D^z U_z + D^\zb U_\zb\right] +\frac{1}{4} N_{zz}N^{zz} + 4\pi \lim_{r\rightarrow \infty} \left[r^2T_{uu}^M\right]}{r^2} + \dots=0,
\eeq
where $(\dots)$ indicates subleading terms in the $1/r$ expansion.  We have used the fact that $T^M_{nU}= T^M_{uu}$ to  leading order in $1/r$ and  $\mathcal{L}_U \Sigma = -n^\mu \Sigma_{,\mu}$. Let
\beq
T_{uu} \equiv \frac{1}{4} N_{zz}N^{zz} + 4\pi \lim_{r\rightarrow \infty} \left[r^2T_{uu}^M\right]
\eeq
represent the injection of energy into the membrane from the outside world.  To leading order, the constraint \eqref{eq:constraint} becomes
\beq\label{eq:constraint2}
\mathcal{M} = \partial_u m_B + \frac{1}{2}\partial_u \left[D^z U_z + D^\zb U_\zb\right] + T_{uu} = 0.
\eeq

Similarly, we can expand the membrane energy equation \eqref{eq:energy} near \im\ using advanced Bondi coordinates \eqref{eq:metricim} and obtain the constraint
\beq\label{eq:constraint3}
\mathcal{M} = \partial_v m_B^- - \frac{1}{2}\partial_v \left[D^z V_z + D^\zb V_\zb \right] - T_{vv} = 0,
\eeq
where $T_{vv}$ represents the loss of energy from the membrane into the outside world.

Plugging \eqref{eq:constraint2}-\eqref{eq:constraint3} into \eqref{eq:M} and integrating by parts gives
\begin{align}\label{eq:M2}
\int_{\mathcal{I}^+} &dx^3 \sqrt{h} \thinspace f  
	\left[\frac{1}{2}\partial_u \left(D^z U_z + D^\zb U_\zb\right) + T_{uu}\right] 
	+ \int_{\mathcal{I}^+_+} dx^2 \sqrt{\gamma}\thinspace f m_B\notag\\
	&= \int_{\mathcal{I}^-} dx^3 \sqrt{h} \thinspace f
	\left[\frac{1}{2}\partial_v \left(D^z V_z + D^\zb V_\zb \right) + T_{vv}\right]
	+\int_{\mathcal{I}^-_-}  dx^2 \sqrt{\gamma}\thinspace f m^-_B,
\end{align}
where \ipp\ denotes the future boundary of \ip\ and \imm\ denotes the past boundary of \im.  We have eliminated terms using the identity
\beq\label{eq:match}
m_B|_{\mathcal{I}^+_-} = m^-_B|_{\mathcal{I}^-_+},
\eeq
which follows from the physically reasonable assumption that the membrane's pressure \eqref{eq:p} is continuous.  Now consider the special case $f=\delta^2(z-w)$ and assume that the surface terms in \eqref{eq:M2} vanish.  Then we obtain
\beq
\int_{\mathcal{I}^+} du  \left[\frac{1}{2}\partial_u \left(\partial_\zb U_z + \partial_z U_\zb\right) + \gamma_{z\zb} T_{uu}\right] 
=\int_{\mathcal{I}^-} dv  \left[\frac{1}{2}\partial_v \left(\partial_\zb V_z + \partial_z V_\zb\right) + \gamma_{z\zb} T_{vv}\right].
\eeq
This is precisely the same as the infinite set of BMS conservation laws at null infinity (compare with (3.12) of \cite{2014JHEP...07..152S}).

\subsection{Physical interpretation}
\label{sec:interp}

The membrane paradigm sheds light on the physical interpretation of BMS conservation laws.  BMS invariance entails ``energy and momentum conservation at every angle.''  Ordinary energy and momentum are not conserved at every angle; outgoing waves need not reach infinity at the same angles as ingoing waves.  Energy and momentum are conserved at every angle only when ordinary energy-momentum is counted together with boundary degrees of freedom.   Boundary degrees of freedom are encoded in the membrane's stress-energy tensor.

The analogy with ordinary fluids is helpful.  Suppose a shear is generated in an ordinary fluid.  The shear is dissipated by viscosity and the fluid's energy density increases.  The combined energy in shearing motions and the fluid's energy density remains constant.  Similarly, the passage of gravitational waves through a null surface generates a shear,
\beq
\sigma_{AB}\sigma^{AB} =\frac{N_{zz} N^{zz}}{2r^2} + \dots.
\eeq
The shear is dissipated by the membrane's shear viscosity, $\eta \sigma^2$, and its energy density, $\Sigma$, increases.  The total energy is conserved.  The difference between the membrane fluid and  ordinary fluids is that, in ordinary fluids, heat conduction tends to redistribute a fluid's energy-density until it becomes uniform.  However, the membrane's energy equation \eqref{eq:energy} has no heat conduction term, so changes to $\Sigma$ remain locked in the location where they first appear (barring further interactions with fields outside the membrane).  As a result, energy is conserved at every angle.

In the language of \cite{2014JHEP...07..152S}, $D^zU_z + D^\zb U_\zb$ is the energy stored in soft gravitons.  Suppose $m_B, N_{zz},$ and $N^{zz}$ revert to zero as $u\rightarrow \infty$.  In this case, the membrane energy density is 
\beq\label{eq:soft}
\Sigma = \frac{1}{r} + \frac{D^zU_z + D^\zb U_\zb}{r^2} + \dots,
\eeq
as $u\rightarrow \infty$.  We see that $\Sigma$ encodes the energy in soft gravitons.

\subsection{Black hole information problem}

Consider an evaporating, one-sided black hole formed from stellar collapse.  The generalized BMS conservation laws,  \eqref{eq:N} and \eqref{eq:M}, imply an infinite set of constraints on \hp\ (take $R=\mathcal{H}^+$ and set the rhs's to zero).  However, they do not give any relationship between \h\ and \i.  So it is not immediately clear that they are useful for the information problem.  Material falling into the black hole imprints its BMS charges on the horizon.  The BMS charges become stored in the membrane's stress-energy tensor (equivalently, soft gravitons on the horizon; see \eqref{eq:soft}).  The problem is that this information can stay on the horizon indefinitely, even as the black hole evaporates and disappears.  A new principle is needed to force the Hawking radiation to carry away the horizon's BMS charges.   The teleological nature of the membrane may be useful in this regard.  The teleological nature of the membrane implies that one must impose future boundary conditions at \hpp, the future boundary of \hp, to determine the evolution of the membrane \cite{1986bhmp.book.....T}.  It is possible that the correct boundary condition at \hpp\ will force the Hawking radiation to carry away the horizon's BMS charges.  This is reminiscent of the black hole final state proposal \cite{2004JHEP...02..008H}.

\subsection{Antipodal matching}
\label{sec:matching}

The distinction between the membrane and the null surface it represents can usually be ignored because the membrane is understood to be arbitrarily close to the null surface.  One case where this distinction is important is for understanding the antipodal identifications of \ip\ and \im\ and of \hp\ and \hm\ \cite{Penrose:1987uia,2014JHEP...07..152S,2015arXiv150206120P,2015arXiv150602906K}.  This point does not seem to have appeared in the membrane paradigm literature, so we discuss it here.

The membrane is a continuous, timelike surface.  The unit normal always points along $\partial_r$ and there is no antipodal identification between different points on the membrane.   The null surfaces are more subtle.  The normal vector, $k$, of a null surface is defined such that $k^\alpha$ is future pointing \cite{poisson2004relativist}.  So $k=2\partial_u=-dv$ on \ip\ and \hm\ and $k=2\partial_v=-du$ on \im\ and \hp. 

\begin{figure}[tbp]
\centering
\includegraphics[width=0.5\textwidth]{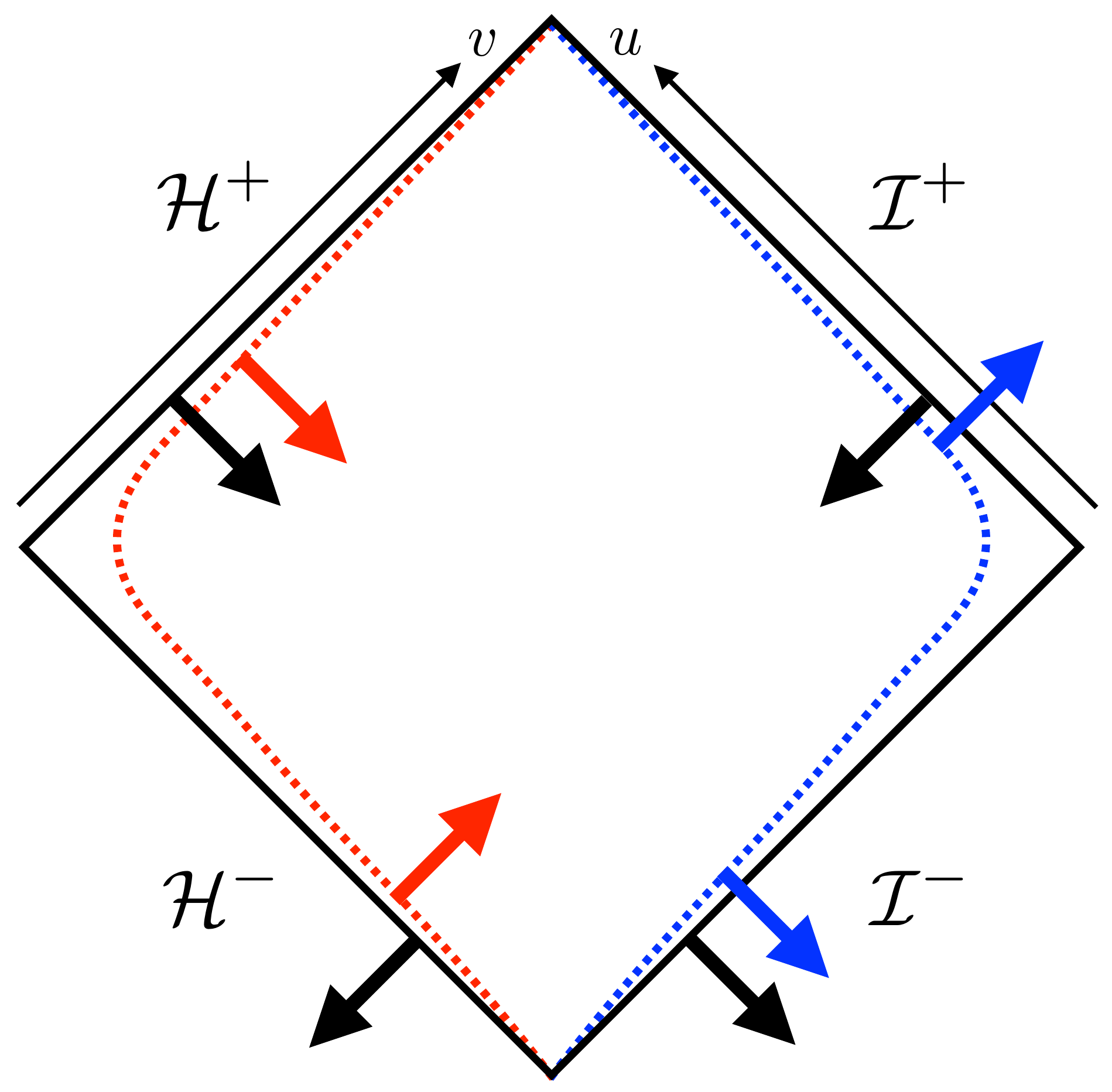}
\caption{Penrose diagram for the exterior of an asymptotically flat, eternal black hole. This is the causal diamond for observers who remain forever in the black hole exterior.  Future and past black hole horizons (\hp\ and \hm) and future and past infinity (\ip\ and \im) are indicated.  The stretched horizon (dotted red) and stretched infinity (dotted blue) are timelike. Heavy colored arrows indicate the unit normals of the membranes.  Heavy black arrows indicate $k_\alpha$, the normals of the null surfaces.}
\label{fig:penrose}
\end{figure}

In the limit as the membrane becomes null, the normals of the membrane and the null surface coincide at \hp\ and \im\ (see Figure \ref{fig:penrose}).  However, at \ip\ and \hm\ they differ by a minus sign.  So the membrane and the null surface are antipodally identified at \ip\ and \hm.  It follows that \ip\ and \im\ are antipodally identified, and \hp\ and \hm\ are antipodally identified.  It is important to keep this in mind when interpreting \eqref{eq:match}: this equation requires an antipodal identification at \i, but not at the membrane.

\section{Charge conservation at every angle}
\label{sec:Q}

Scattering in electrodynamics is governed by an infinite dimensional symmetry entailing charge conservation at every angle \cite{2014arXiv1412.2763K,2015arXiv150302663H,2015arXiv150500716P,2015arXiv150602906K,Susskind:2015hpa}.  Charge conservation at every angle appears in the membrane paradigm as the charge continuity equation on the membrane \cite{1986bhmp.book.....T,1998PhRvD..58f4011P,2015PhRvD..91h4044P},
\beq\label{eq:charge}
\frac{\partial \sigma}{\partial \lambda} + \vec{\nabla}_{(2)} \cdot \vec{j} = -J^n. 
\eeq
where
\begin{align}
\sigma &= \pm l_a n_b F^{ab},\\
(\vec{j} \times \hat{n})^a  &= \pm \epsilon^a_b h^b_i n_j F^{ij},\\
J^n &= J^a n_a,
\end{align}
The upper sign is taken on the membrane at infinity and the lower sign is taken on the membrane at the horizon. $\sigma$ is the membrane charge density.  It terminates the normal component of the external electric field at the membranes.   $\vec{j}$ is the membrane current density.  It terminates the transverse component of the external magnetic field at the membranes.  $J^n$ represents charges falling into the membranes from the external universe.  Eq. \eqref{eq:charge} expresses the fact that charges falling into the membranes are captured by the membrane charge density and currents.  The total charge  is conserved at each point on the membrane.

To see the equivalence of \eqref{eq:charge} with the infinite number of charge conservation laws discussed by \cite{2015arXiv150602906K}, consider the expansion of \eqref{eq:charge} near \ip.  In this case, 
\begin{align}
\partial_\lambda \sigma &= \partial_u F_{ru},\\
 \vec{\nabla}_{(2)} \cdot \vec{j} &= \partial_u (D^z A_z + D^\zb A_\zb).
\end{align}
Now multiplying \eqref{eq:charge} by an arbitrary function $\epsilon(z,\zb)$ and integrating over all of \ip\ gives a charge, $Q^+_\epsilon$.  Similarly, we may expand \eqref{eq:charge} near \im, multiply by  $\epsilon(z,\zb)$, and obtain a charge, $Q^-_{\epsilon}$.    On stretched infinity, $Q^+_\epsilon=Q^-_\epsilon$ because stretched infinity is a single, continuous fluid.  On true infinity we enforce the same conservation laws by  assuming the matching conditions discussed in section \ref{sec:matching}.  

\subsection{Lienard-Wiechert fields}

To illustrate the difference between stretched infinity and true infinity, consider the field of a point charge moving with constant velocity $\beta$.  We assume the charge passes through the origin at $t=0$. In this case,
\begin{align}
E_r &= \frac{q}{4\pi}\frac{\gamma(r-\beta t\cos\theta )}{(\gamma^2(t-\beta r\cos\theta)^2-t^2+r^2)^{3/2}}, \\ 
B_\phi &= \frac{q}{4\pi}\frac{\gamma\beta r\sin\theta}{(\gamma^2(t-\beta r\cos\theta)^2-t^2+r^2)^{3/2}},
\end{align}
where $\gamma^2 = 1/(1-\beta^2)$.  The field satisfies Ampere's law
\beq\label{eq:ampere}
\frac{\partial E_r}{\partial t} - \frac{1}{r\sin\theta} \frac{\partial}{\partial\theta} (\sin\theta B_\phi) = 0.
\eeq
This is equivalent to the membrane charge conservation law \eqref{eq:charge} on stretched infinity upon making the identifications $\sigma = -E_r$ and $j_\theta = B_{\phi}$.   

Setting $t=u+r$ and expanding near $r=\infty$ gives
\beq
\sigma_+ = -\frac{q}{4\pi r^2}\frac{1}{\gamma^2(1-\beta\cos\theta)^2} + \dots.
\eeq
Setting $t=v-r$ and expanding near $r=\infty$ gives
\beq
\sigma_- = -\frac{q}{4\pi r^2}\frac{1}{\gamma^2(1+\beta\cos\theta)^2} + \dots.
\eeq
$\sigma_+$ and $\sigma_-$ differ by a minus sign in the denominator.  The interpretation of this sign flip is different on stretched infinity and true infinity.  
On true infinity, \im\ and \ip\ are antipodally identified: $\theta \rightarrow \pi-\theta$ (see section \ref{sec:matching}).  So $\sigma_-=\sigma_+$ and charge is conserved at every angle because the charge density is constant (there is no current flow on the membrane).

On stretched infinity, the $S^2$'s in the far past and the far future are not antipodally identified.  So on stretched infinity $\sigma_-$ must evolve into $\sigma_+$.  According to eq. \eqref{eq:ampere}, membrane currents flow on stretched infinity in just the right way so as to convert $\sigma_-$ in the far past into $\sigma_+$ in the far future.  

As stretched infinity approaches true infinity, the membrane current flow becomes concentrated near spatial infinity, $i_0$.  In the true infinity limit, the membrane current flow disappears but charge conservation at every angle is preserved by the antipodal identification between \im\ and \ip.

\subsection{Symmetries}
\label{sec:symmetries}

Ordinary charge conservation follows from global $U(1)$ invariance.  That is, we have a theory with a phase, $\psi$, which is invariant under constant shifts,
\beq\label{eq:u1}
\psi \rightarrow \psi +c.
\eeq
In a fluid with negligible diffusion, the charge in each fluid element is separately conserved.  In this case,  global $U(1)$ symmetry is enhanced to an infinite dimensional symmetry \cite{2012PhRvD..85h5029D}
\beq
\psi \rightarrow \psi + c(\phi^I),
\eeq
where $c(\phi^I)$ is now an arbitrary function of $\phi^I$, a coordinate labeling the fluid elements.  This enhanced symmetry implies the local charge conservation law \eqref{eq:charge} \cite{2012PhRvD..85h5029D}.  This is the same infinite dimensional symmetry that appears in the description of scattering in electrodynamics in asymptotically flat spacetimes as ``large gauge transformations.''  So again there is a connection between the fluid and gravity pictures at the level of symmetries.

\acknowledgments

I thank Glenn Barnich, Vyacheslav Lysov, Achilleas Porfyriadis, and Andrew Strominger for comments. This work was supported by a Pappalardo Fellowship in Physics at MIT.

\bibliography{ms}

\end{document}